**Electrorheological study of the nematic LC 4-n-hepthyl-4'-cyanobiphenyl: experimental and theoretical treatment**

**M.T. Cidade[1], C.R. Leal[1,2] and P. Patrício[2,3]**

[1]Materials Science Department and CENIMAT/I3N, New University of Lisbon, Campus da Caparica, 2829-516 Caparica, Portugal
[2]ISEL, Polytechnical Institut of Lisbon, Scientific Area of Physics, Rua Conselheiro Emídio Navarro,1,1949-014 Lisboa, Portugal
[3]C.F.T.C.-Lisbon University, Av. Prof. Gama Pinto 2, P-1649-003 Lisboa, Portugal

**Abstract:** The experimental and theoretical study of the electrorheological (ER) effect observed in the nematic phase of the 4-n-hepthyl-4'-cyanobiphenyl (K21) is the aim of this work. The K21 liquid crystal (LC) appears to be a model system where all the observed rheologial behaviours can be interpreted by the Leslie-Ericksen (L-E) continuum theory for low molecular weight liquid crystals. We present the flow curves of our sample for different temperatures and under the influence of an external electric field, ranging from 0 to 3kV/mm, applied perpendicular to the flow direction. We also present the viscosity as a function of the temperature, for the same values of electric field, obtained for different shear rates. A master flow curve was built, dividing the shear rate by the square of the electric field and multiplying by the square of a reference electric field value, for each temperature, where two Newtonian plateaus appear at low and high shear rate values, connected by a shear-thinning region at intermediate shear rate values. Theoretical interpretation of the observed behaviours is proposed in the framework of the continuum theory. In this description the director alignment angle is a function of the electric field and the flow field - boundary conditions are neglected. In this way it was possible to extract some viscoelastic parameters, as well as the dielectric anisotropy.

**INTRODUCTION**

The electrorheological (ER) effect can be considered as the change of the fluid apparent viscosity in the presence of an external electric field [Wen et al. (2008)], sometimes having dramatic consequences, inducing the fluid solidification [Wen *et al.* (2008); Larson (1999)]. The ER effect is an important phenomenon that brings new technological applications which may be used in a wide spectrum of domains, spanning from the electro-optical device to automobile industries. Think for example on shock absorbers made out of tunable vibration damping systems.

The first observation of the ER effect was reported by Winslow (1949), for a suspension of polarizable micro particles adsorbed in water. In this case, the variation of the apparent viscosity was due to the micro particles alignment in chains or columnar structures oriented in the direction of the electric field, increasing the solution's viscosity in one order of magnitude.

Liquid crystalline materials are known to exhibit an ER effect, since they are susceptible to be oriented by external electric and magnetic fields [de Gennes (1993)] in their anisotropic phases.

In the last two decades, several research articles were devoted to the experimental characterization of the ER effect in different liquid crystalline systems, some of which proposed theoretical quantitative interpretation of the experimental results.

The ER effect was studied in lyotropic liquid crystalline polymers such as poly(n-hexyl isocianate) in p-xylene (PHIC/p-xylene) [Yang and Shine (1992)], in poly(γ-benzyl-L-glutamate) in 1,4-dioxane (PBLG/1,4-dioxane) [Tanaka *et al.* (1997); Neves *et al.* (2008)] and in acetoxypropylcellulose in dimethylacetamide (APC/DMAc) [Neves *et al.* (2008)]. A significant increase (about four times for an electric field of 2kV/mm) of the viscosity was observed for solutions of PBLG/1,4-dioxane [Tanaka *et al.* (1997); Neves *et al.* (2008)]. Inversely, a decrease of the viscosity was measured for APC/DMAc solutions, since this system has a negative dielectric susceptibility.

In thermotropic liquid crystalline polymers, like polysiloxanes based liquid crystal (LC) polymers [Inoue and Maniwa (1995); Inoue *et al.* (1997); Kaneko et al. (2008)], the ER effect was also reported. Here, a significant increase (of about one order of magnitude) of the viscosity has been observed due to the orientation effect of side-chain mesogenic groups attached to the polymeric backbone.

For low molecular weight liquid crystals, several systems have also been studied, and most of them were carried out by K. Negita [(1995); (1996); (1997); (1999)]. Starting with the liquid crystal MBBA, the ER effect was studied in the nematic phase. For this system, a small decrease of the viscosity with the applied electric field was observed, which is justified by the negative dielectric susceptibility of MBBA [Negita (1995)]. The influence of different liquid crystalline phases - namely smetic A and nematic phases - in the ER effect, was also considered in the experiments performed with 4-n-octyl-4'-cyanobiphenyl (8CB) [Negita (1997)] and 4-n-octyloxy-4-cyanobiphenyl (8OCB) [Negita (1999)]. From these studies, it is clear that in the nematic phase, a significant increase of the viscosity is observed in the presence of the electric field, and the opposite effect is obtained for the smectic A phase. In this system, the electric field re-orients the smectic layers and the nematic director, causing the ER effect. A thorough understanding of the ER effect in the nematic phase was presented for the 4-n-pentyl-4'-cyanobiphenyl (5CB) [Negita (1996)], where important properties were clearly evidenced: (i) the ER effect – in this case, the viscosity increase under the application of the electric field - is observed in the nematic phase but not in the isotropic phase; (ii) the flow changes its behaviour from a Newtonian to a non-Newtonian and again to a Newtonian fluid, as the electric field is increased; (iii) the viscosity increase saturates at high enough electric fields; (iv) the ER effect is dependent on shear rate and temperature.

It was also established [Negita (1995); (1996); (1997); (1999)] that in liquid crystalline systems, in the nematic phase, the ER effect is caused by the reorientation of the nematic director, which is determined by a balance between the flow field and the electric field. When the applied electric field is perpendicular to the flow direction, the apparent viscosity increases when the system has a positive dielectric anisotropy (meaning the molecules prefer to be aligned along the electric field). When the dielectric anisotropy of the system is negative the opposite behaviour is induced and a decrease in the apparent viscosity is observed.

In this article, a first systematic experimental characterization of the viscosity of the liquid crystal 4-n-hepthyl-4'-cyanobiphenyl (K21) is presented, as a function of the shear rate, the temperature, and the applied electric field. A theoretical interpretation of the observed behaviours is proposed in the framework of the continuum theory of Leslie-Ericksen (L-E) for low molecular weight nematic liquid crystals [Leslie (1966); Ericksen (1960)].

It is our believe that a *model* system was found, since all the experimental rheological behaviours can be accounted by the L-E theory. In particular, we show that the flow curves obtained with and without the application of an electric field, for different intensity values, fall into a viscosity master curve that describes the total expected behaviour. Other interesting phenomenon is the viscosity vs temperature behaviours, in particular in the nematic-isotropic transition region, where a continuous dependence for the viscosity with temperature is observed, independently of the applied shear rate and applied electric field values. The ER effect vanishes above the nematic-isotropic transition temperature, where the nematic order is no longer defined.

**EXPERIMENTAL**

The liquid crystal used in this study, the 4-n-hepthyl-4´-cyanobiphenyl, is commercially known by K21 (Merck). Its molecular structure is presented in figure 1. It presents a nematic phase between 30 and 42.8 ºC.

We have studied the electrorheological properties of K21 using a Bohlin Gemini HR$^{nano}$ rotational rheometer, to which a Bohlin electrorheological cell was coupled. The geometry used was plate/plate, with a gap of 500 μm. After temperature equilibration, and before starting the measurements, the samples were subjected to a pre-shearing stage, with a pre-shear of 10 s$^{-1}$ applied for 60 s, followed by an equilibration time of 120 s.

We have measured the viscosity in function of the shear rate for different temperatures in the nematic phase, without and with the application of an external electric field in the perpendicular direction with respect to the flow direction, for electrical fields in the range

0 – 3 kV/mm. We have also measured the viscosity in function of the temperature covering the nematic and the isotropic phase, for different shear rates and different values of the applied electric field.

**EXPERIMENTAL RESULTS**

**Flow curves – ER effect**

Figure 2 presents the flow curves of K21 for different electric fields applied, at $T = 32$ and 36 ºC, as examples. Curves for other temperatures were also performed and show similar trends, even though, as expected, differences in the absolute values of the viscosity were found.

The analysis of figure 2 shows that the liquid crystal, in the absence of an electric field, presents a Newtonian behaviour. The application of an electric field increases the viscosity in the low and intermediate shear rate ranges, while for the higher shear rates the curves obtained when the sample is under an electric field tends to converge with the curve obtained without electric field applied.

The form of the flow curve is due to the competition between the electric and the flow fields. In fact, the molecule's orientation of a liquid crystal of positive dielectric anisotropy, which is the case of K21, will follow the direction of the applied electric field. As shown in figure 3, if the electric field is applied perpendicular to the flow field, the molecules will re-orient themselves in the direction of the electric field, or perpendicular to the flow field, as long as the flow field is not strong enough to avoid it, thus increasing the sample's viscosity. For high flow field strengths, which means high shear rates, the flow field is known to orient the molecules of a liquid crystal in a direction that makes a relatively small angle with the flow field, the flow-alignment angle, $\theta_0$ [Larson (1999)]. This is in fact what happens with our sample. For sufficiently high flow field strengths, the flow field dominates over the electric field and the molecules orient in such a direction that its viscosity decreases and all the curves tend to the curve without electric field applied. Of course, the shear rate in which a strong decrease of the viscosity is observed depends on the electric field strength, which means that for higher electric fields applied, higher shear rates are necessary to cancel the effect of the electric field. For intermediate shear rates, a competition between both fields exist, and the molecules orient in a direction that makes an angle $\theta$ with the flow field, angle that depends on the electric field (it increases with the electric field strength).

Figure 2 also shows that, for the lower shear rates of our experiment, a critical electric field of 0.26 kV/mm exists for this system. Until this value of electric field, an increase in the viscosity, for low shear rates, is observed when increasing the electric field. However, the increase of the electric field value above 0.26 kV/mm does not lead to a further increase in the viscosity, which means that an electric field of 0.26 kV/mm is sufficient to completely orient the molecules along with it.

**Viscosity vs Temperature – ER effect**

Figure 4 presents the influence of the temperature in the viscosity of K21 without and with electric fields up to 3 kV/mm applied, for shear rate values of 10 s$^{-1}$ and 100 s$^{-1}$. We obtained the same curves for shear rates of 50 and 500 s$^{-1}$, which presented similar behaviours, except for the absolute values of the viscosity, as expected.

The analysis of figure 4 shows that the isotropic phase, as expected, is not influenced by the electric field. That is why, at 43 ºC, all the curves converge to one single curve, the one presented without any electric field applied. This region of the curve shows a slight decrease of the viscosity with increasing temperature. We also observe a slight decrease in the viscosity with the increase of the temperature in the nematic phase, far away from the clarification temperature. In the vicinity of the nematic-isotropic transition temperature different behaviours were observed, depending on the value of the electric field. Without electric field, or even with very small values of electric field, a sudden increase of the viscosity is observed at the transition, which is the usual behaviour of liquid crystals and it is due to the fact that in the nematic phase the molecules present an orientational order (characterized by a positive order parameter), which facilitates the flow, while in the isotropic phase the order parameter is zero. However, for higher values of the applied electric field, when there is an already sufficiently high enhancement of the viscosity, the behaviour is totally altered and, at the transition from the nematic to the isotropic phase a decrease in the viscosity is observed. Once again it is the competition between both electric and flow fields that settles the behaviour, and the temperature at which the slope of the $\eta(T)$ curve significantly increases depends on the electric field strength.

**Flow master curve**

The building of the viscosity master curve starts with the definition of a reference electric field, for which we have chosen the critical electric field of 0.26 kV/mm, above which no enhancement of the viscosity is induced in the system by the increasing of the applied electric field (for the lower shear rates of our experiment).

Figure 5 presents the master curve of viscosity vs shear rate, for T = 32ºC, which we have obtained by dividing the shear rate by the square of the electric field and multiplying it by the square of the reference electric field (the critical electric field of 0.26 kV/mm).

Fitting the master curve data to the L-E theory allowed us to extrapolate some viscoelastic parameters and the dielectric anisotropy, as shown in the next section.

**THEORETICAL RESULTS**

**Leslie-Ericksen theory**

Leslie-Ericksen (L-E) theory [Leslie (1966); Ericksen (1960)] describes the dynamical behaviour of a low-molecular nematic liquid-crystal. It generalizes the Navier-Stokes theory of simple fluids, coupling the velocity field of the molecules with their average orientation, which is given by the nematic director field, **n**. The L-E theory takes into consideration the Frank-Oseen free elastic energy $f_F$, which is a function of the nematic distortions [Larson (1999)]:

$$2f_F = K_1(\nabla \cdot \mathbf{n})^2 + K_2(\mathbf{n} \cdot \nabla \times \mathbf{n})^2 + K_3(\mathbf{n} \times \nabla \times \mathbf{n})^2 - \epsilon_a(\mathbf{n} \cdot \mathbf{E})^2 \quad (1)$$

where $K_1$, $K_2$ and $K_3$ are the splay, twist and bending Frank elastic constants. The molecule's tendency to align parallel (perpendicular) to the electric field **E** is given by the positive (negative) sign of the dielectric anisotropy $\epsilon_a$. Using Einstein's notation for summation of indices, we may write

$$2f_F = K_1(\partial_i n_i)^2 + K_2(\epsilon_{ijk} n_i \partial_j n_k)^2 + K_3(n_j \partial_i n_j)(n_k \partial_i n_k) - \epsilon_a(n_i E_i)^2 \quad (2)$$

where $\epsilon_{ijk}$ is the antisymmetric tensor. The dynamical equations are

$$\partial_i v_i = 0 \quad (3)$$

$$\rho \frac{dv_i}{dt} = \partial_j \sigma_{ji} \quad (4)$$

$$\sigma \frac{d^2 n_i}{dt^2} = g_i \quad (5)$$

The first equation expresses the conservation of mass, and **v** is the velocity field. The second equation corresponds to Newton's law of motion for the particles of the fluid. $\rho$ is the mass density, and $\tilde{\sigma}$ is the stress tensor. The third equation expresses the motion of the director orientation and $\sigma$ is an inertial coefficient. The vector **g** and the stress tensor are defined by the constitutive equations, which are usually written in the form:

$$\sigma_{ij} = -p\delta_{ij} + \sigma_{ij}^d + \sigma_{ij}' \tag{6}$$

where $p$ is the pressure. The Frank distortional stress is given by

$$\sigma_{ij}^d = -\left(\frac{\partial f_F}{\partial(\partial_i n_k)}\right)\partial_j n_k \tag{7}$$

and the viscous stress is

$$\sigma_{ij}' = \alpha_1 n_i n_j n_k n_l V_{kl}^s + \alpha_2 n_i N_j + \alpha_3 n_j N_i + \alpha_4 V_{ij}^s + \alpha_5 n_i n_k V_{kj}^s + \alpha_6 n_j n_k V_{ki}^s \tag{8}$$

where

$$V_{ij}^s = \frac{1}{2}(\partial_i v_j + \partial_j v_i) \tag{9}$$

is the symmetrical part of the velocity gradient tensor and

$$N_i = \frac{dn_i}{dt} - \epsilon_{ijk} w_j n_k \tag{10}$$

is the rotation rate of **n** relative to that of the background fluid, expressed by the angular velocity $w_i = \epsilon_{ijk}\partial_j v_k /2$. $\alpha_1,...,\alpha_6$ are the L-E viscosity coefficients. Finally, the field **g** is given by

$$g_i = \lambda n_i - \frac{\partial f_F}{\partial n_i} + \partial_j\left(\frac{\partial f_F}{\partial(\partial_j n_i)}\right) - \gamma_1 N_i - \gamma_2 n_j V_{ji}^s \tag{11}$$

The viscosity coefficients $\gamma_1 = \alpha_3 - \alpha_2$ and $\gamma_2 = \alpha_3 + \alpha_2$. $\lambda$ is a Lagrange multiplier to ensure $n_i n_i = 1$. The L-E coefficients are not independent, but they must satisfy Parodi's equation $\alpha_6 = \alpha_2 + \alpha_3 + \alpha_5$. If the director is held in a fixed orientation by an electric field strong enough to overcome the effects of the flow, then shear-rate independent viscosities can be measured. The three simplest of these are called the Miesowicz viscosities (see figure 3): $\eta_a = (-\alpha_2 + \alpha_4 + \alpha_5)/2$; $\eta_b = (\alpha_3 + \alpha_4 + \alpha_6)/2 = \eta_a + (\alpha_2 + \alpha_3)$ and $\eta_c = \alpha_4/2$.

In our experiment, the rheometer performs a simple horizontal Couette flow, $\mathbf{v} = (v,0,0)$, with an applied perpendicular electric field, $\mathbf{E} = (0,0,E)$ (see figure 3). The thickness of our cell is large enough to neglect the effect of the nematic anchoring conditions at the surfaces. It was verified that the wall effect should be noticeable only for a gap inferior to 50 μm. Then, the bulk orientation of the nematic is constant, $\mathbf{n} = (\cos\theta, 0, \sin\theta)$, independent of space and time, and the velocity gradient tensor will have only one component, the constant shear rate $\partial_z v_x = \dot{\gamma}$. With these assumptions, the mass conservation and Newton's law (equations (3) and (4)) are automatically satisfied. From equation (5), we have

$$(\alpha_3 \cos^2\theta - \alpha_2 \sin^2\theta)\dot{\gamma} - \frac{1}{2}\epsilon_a E^2 \sin 2\theta = 0 \tag{12}$$

This condition determines the value of the angle $\theta$ as a function of $\dot{\gamma}/(\epsilon_a E^2)$ for a given set of fluid viscosities $\alpha_2$, and $\alpha_3$. This shows why all flow curves, for different applied electric fields, may be plotted into the same master curve, only redefining the scale of the shear rate by $\dot{\gamma} E_{ref}^2 / E^2$, where $E_{ref}$ is a reference electric field (see figure 5).

The viscosity, $\eta = \sigma_{zx}/\dot{\gamma}$, can be obtained from the stress tensor (equation (6)) and depends on the angle $\theta$, and the viscosity coefficients of the fluid:

$$\eta = \alpha_1 \sin^2\theta \cos^2\theta + \eta_b - (\alpha_2 + \alpha_3)\sin^2\theta \tag{13}$$

For sufficiently large $E$, the solution given by equation (12) is $\theta_{max} = \pi/2$, and the director will be parallel to the electric field, if $\epsilon_a > 0$. In this case, the viscosity is $\eta_{max} = \eta_b - (\alpha_2 + \alpha_3) = \eta_a$. When no electric field is applied, the solution of equation (12) is $\theta_{min} = \arctan\sqrt{\alpha_3/\alpha_2}$. The viscosity becomes $\eta_{min} = \alpha_1\alpha_2\alpha_3/(\alpha_2+\alpha_3)^2 + \eta_b - \alpha_3$.

We have used the Levenberg-Marquardt method [Press (1992)] to fit the flow master curve (figure 5), for the temperatures 32º, 34º, 36º and 38º C. For every value of $\dot{\gamma} E_{ref}^2 / E^2$, we have obtained $\theta$ from equation (12). The viscosity was then calculated from equation (13), with a total of 5 nonlinear fitting parameters: $\alpha_1$, $\alpha_2$, $\alpha_3$, $\eta_b$, and $\epsilon_a$. We have noticed that our fits were very sensitive to the experimental accuracy. We have verified that it was possible to neglect $\alpha_1$ (which also happens for other materials [e.g. Larson (1999); Negita (1997)], especially when we are close to the nematic-isotropic transition). We have also verified that a considerable change in $\alpha_3$ (from −10 to −30 mPa.s) almost did not change the quantitative behaviour of the viscosity, as long as $\eta_b$ varied by the same amount, meaning these two parameters are not completely independent in our fit. So, we have redone the fits with only three simple parameters, $\eta_{min} = \eta_b - \alpha_3$, $\eta_{max} - \eta_{min} = -\alpha_2$ and $\epsilon_a$, which may be easily and almost directly estimated from the experimental data. We have obtained these parameters also from the experimental viscosity-temperature curves (figure 4 and related), using the re-dimensioned shear rates $\dot{\gamma} E_{ref}^2 / E^2$. This allowed us to extend our fits for other temperatures, in between 31º and 42º C (slightly below the nematic-isotropic transition), in spite of a worse accuracy, due not only to the experimental results, but also introduced by the numerical procedure.

From the analysis of figure 6 (a) and (b), it is possible to see that $\alpha_2 \approx -80$ mPa.s at 32º C, and increases rapidly, almost linearly to zero, as we approach the nematic-isotropic temperature. Also, $\eta_{min} = \eta_b - \alpha_3 \approx 23$ mPa.s at 32º C, and decreases slowly to a constant 18 mPa.s at the transition – the L-E theory predicts indeed that $\alpha_2$ and $\alpha_3$ vanish, and all the Miesowicz viscosities tend to only one isotropic viscosity at the nematic-isotropic transition. The dielectric anisotropy $\epsilon_a$ is represented in figure 7. The points are scattered, but it is still possible to observe a slight linear decrease, from $6{,}2\epsilon_0$ ($\epsilon_0$ is the permittivity of vacuum) at 32º C to $5{,}9\epsilon_0$ at 40ºC. The dielectric anisotropy can be used as an order parameter of the nematic phase. It is proportional to the degree of alignment of the molecules and it is expected to decrease rapidly to zero close to the transition [Larson (1999), pp.456-457].

**CONCLUSIONS**

We have studied the influence of the application of an electric field, perpendicular to the flow field, in the rheological behaviour of LC 4-n-hepthyl-4'-cyanobiphenyl (K21). Our experimental data lead us to the following conclusions:

(i) 4-n-hepthyl-4'-cyanobiphenyl presents an ER effect, with an increase of about five times the viscosity for small shear rates and moderate electric field strengths.
(ii) For the lower shear rates of our experiment, a critical electric field of 0.26 kV/mm exists for this system. Until this value of electric field, an increase in the viscosity, for low shear rates, is observed when increasing the electric field. However, the increase of the electric field value above 0.26 kV/mm does not lead to a further increase in the viscosity, which means that an electric field of 0.26 kV/mm is sufficient to completely orient the molecules along with it.
(iii) For increasingly higher shear rates a competition between the flow field and the electric field occurs and we observe a decrease in the viscosity enhancement.

(iv) For sufficiently high shear rates the flow field becomes dominant and the flow curves obtained with different electric fields applied converge to the curve obtained without electric field.
(v) In the nematic phase we observe a slight decrease in the viscosity with the increase of the temperature, however, near the nematic isotropic transition temperature, a completely different behaviour is observed, depending on the value of the electric field. Without electric field or with small electric fields applied, an increase of the viscosity is observed (which is the usual behaviour for liquid crystals without any electric field applied, and that is due to the molecular orientation in the nematic phase which is no longer present in the isotropic phase), while for higher electric field strengths an abrupt decrease in the viscosity is observed. For temperatures higher than $T_{NI}$ all curves converge to the one without electric field, which means that, as expected, the isotropic phase is not oriented by the electric field.
(vi) We were able to build a master flow curve, after flow curves obtained for different temperatures, by dividing the shear rate by the square of the electric field and multiplying it by the square of a reference electric field.

By fitting the Leslie – Eriksen Theory to our data, we were able to extract some conclusions pertaining viscoelastic parameters and the dielectric anisotropy:

(vii) It was possible to neglect $\alpha_1$.
(viii) At T = 32 ºC, $\alpha_2 = -80$ mPa.s, and it increases, almost linearly, to zero, as $T_{NI}$ is approached.
(ix) At T = 32 ºC, $\eta_{min} = \eta_b - \alpha_3 = 23$ mPa.s, and it decreases slowly to a constant value of 18 mPa.s at $T_{NI}$.
(x) We have observed a slight linear decrease in the dielectric anisotropy, from $6.2\,\epsilon_0$ ($\epsilon_0$ is the permittivity of vacuum), at T = 32 ºC, to $5.9\,\epsilon_0$ at T = 40 ºC.

These findings are all accounted for in the Leslie-Ericksen Theory. In fact, our system turns out to be a *model* system for the L-E theory, that describes well all the rheological behaviours observed. In particular a master flow curve can be built, where the influences of shear rate and electric field are systematically presented.

**Acknowledgements:**
The authors are thankful to Professor João Paulo Casquilho for fruitful discussions.

**Figure Captions**

Figure 1. Molecular formula of 4-n-hepthyl-4'-cyanobiphenyl (K21) and its transition temperatures

Figure 2. Flow curves (viscosity vs shear rate) for different values of electric field at 32 ºC (a) and 36 ºC (b), respectively.

Figure 3. Schematic representation of the molecular orientation for different flow field strengths.

Figure 4. Viscosity in function of temperature (for different values of electric field) for shear rate values of 10 s$^{-1}$ (a) and 100 s$^{-1}$ (b), respectively.

Figure 5. Master curve of the viscosity vs scaled shear rate, at 32ºC. The normalization was performed by dividing the shear rate by the square of the electric field and multiplying it by the critical electric field of 0.26 kV/mm.

Figure 6. Viscosities $\alpha_2$ (a) and $\eta_{min} = \eta_b - \alpha_3$ (b) as functions of temperature. The symbols (X) correspond to the flow master curves fits, performed at 32º, 34º, 36º and 38º C. The symbols (+) correspond to the fits calculated from the viscosity-temperature figures. The tendency line is represented in continuous.

Figure 7. Dielectric anisotropy $\epsilon_a$ (in units of the permittivity of vacuum $\epsilon_0$) as a function of temperature. The symbols (+) correspond to the fits calculated from the viscosity-temperature data (figure 4). The tendency line is represented in continuous.

Figure 1

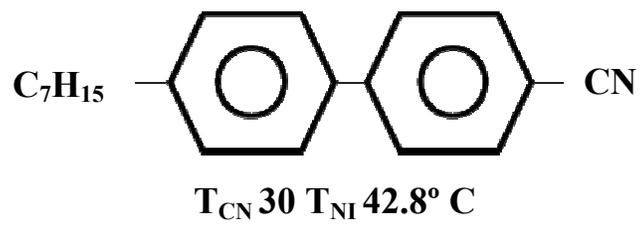

$T_{CN}$ 30 $T_{NI}$ 42.8° C

Figure 2

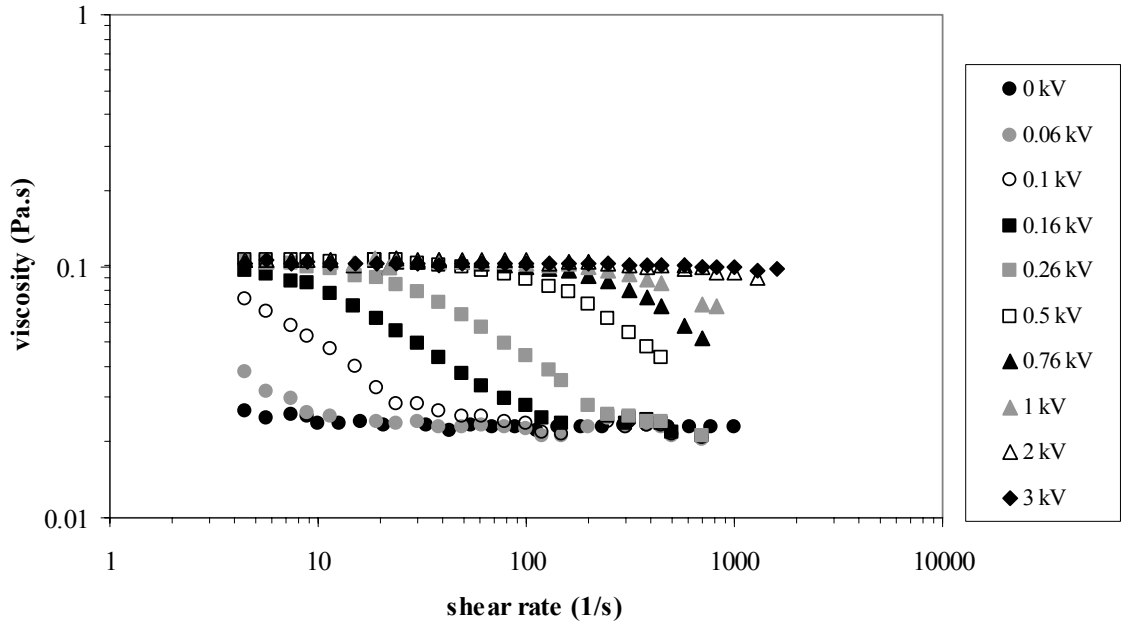

(a)

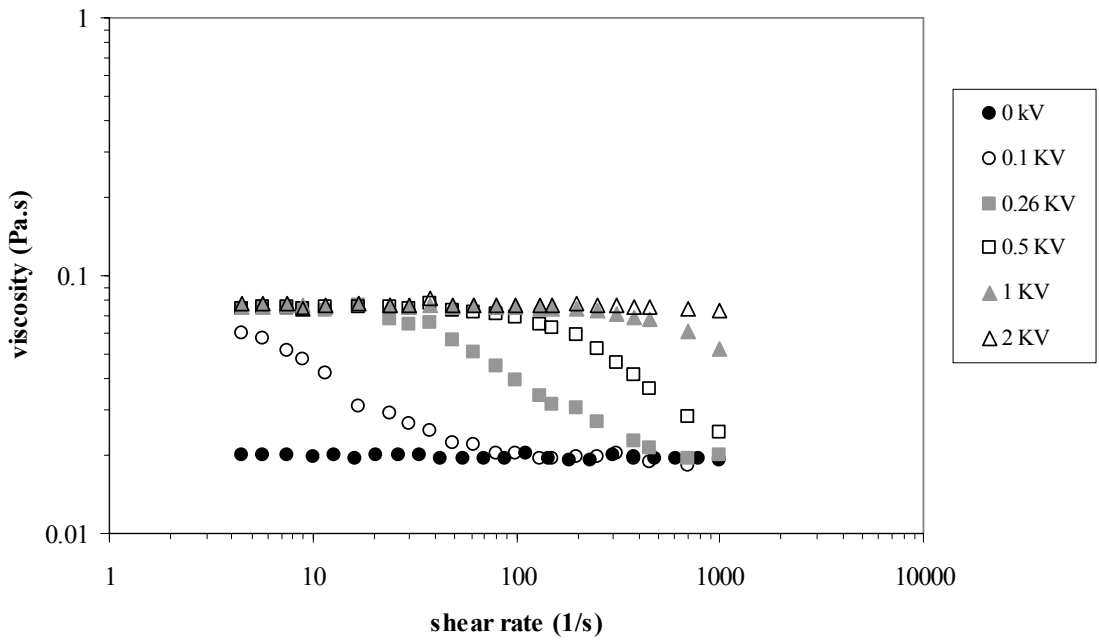

(b)

Figure 3

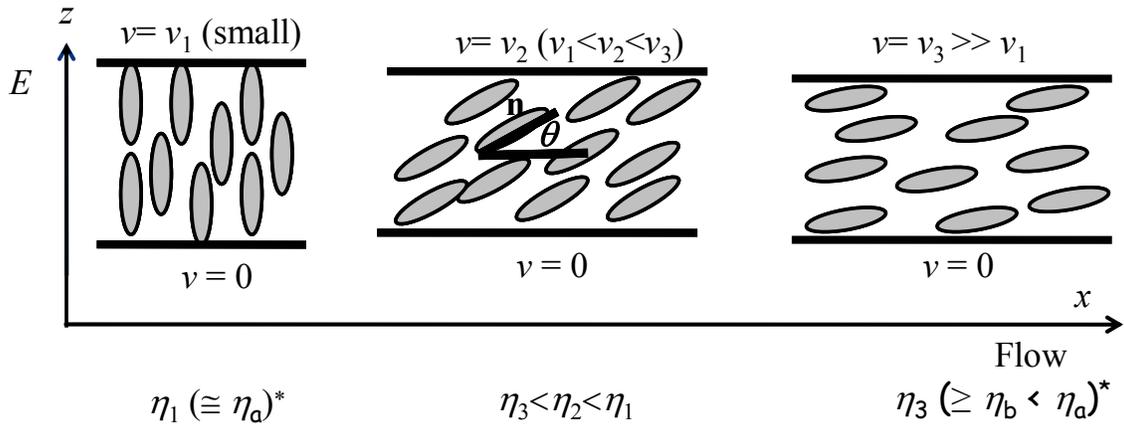

$\eta_1 \;(\cong \eta_a)^*$     $\eta_3 < \eta_2 < \eta_1$     $\eta_3 \;(\geq \eta_b < \eta_a)^*$

*$\eta_a$, $\eta_b$ and $\eta_c$ : Miesowicz viscosities

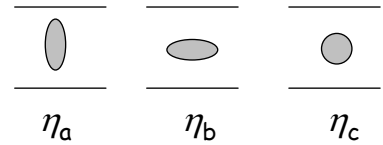

Figure 4

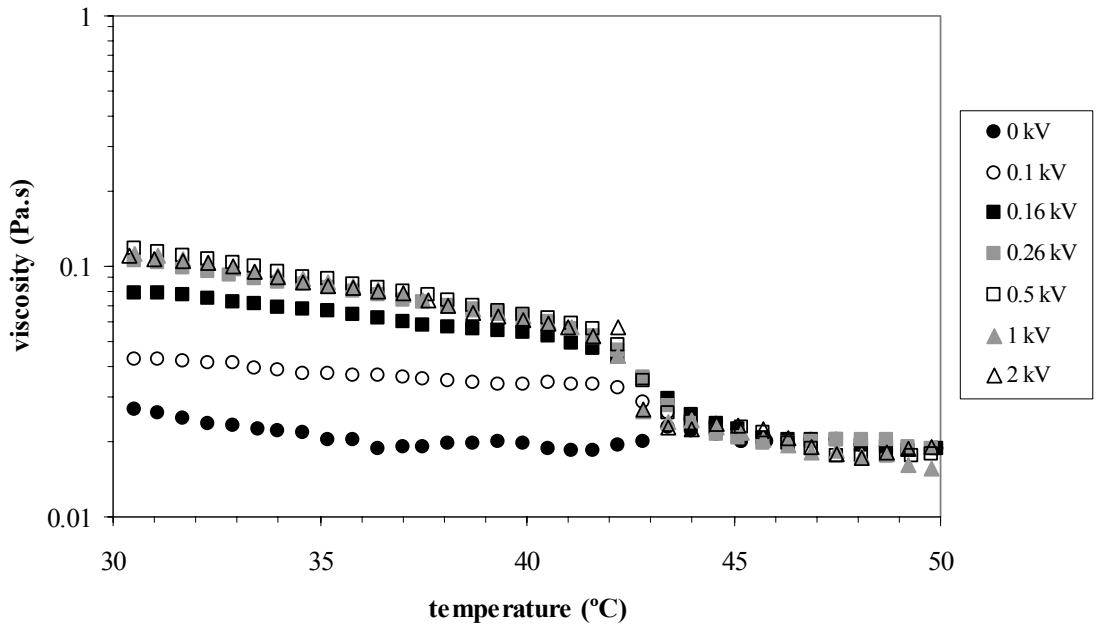

(a)

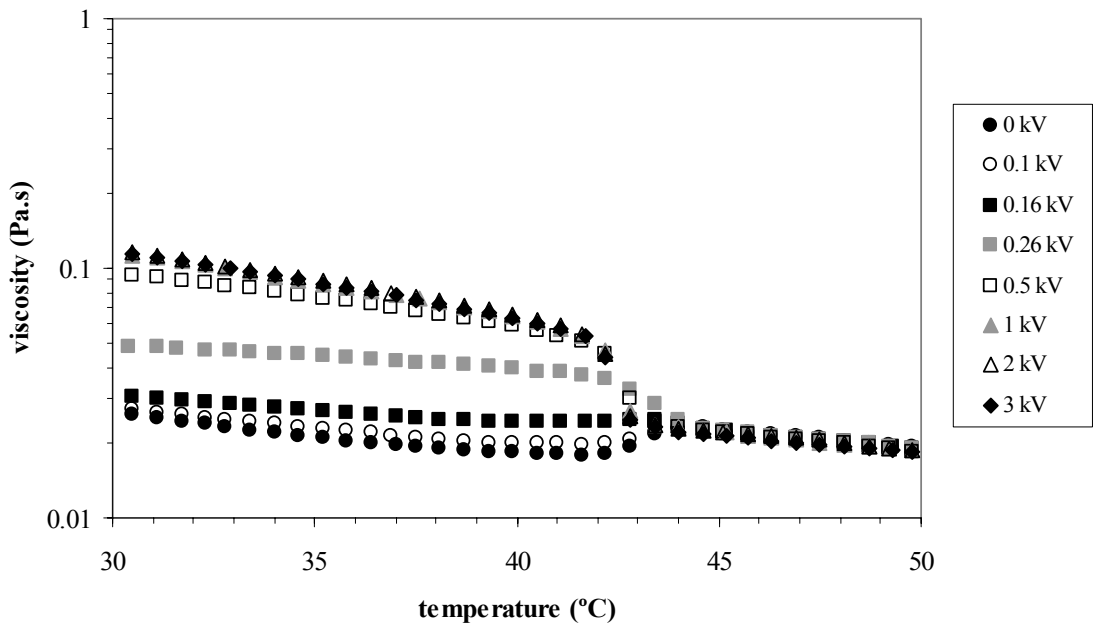

(b)

Figure 5

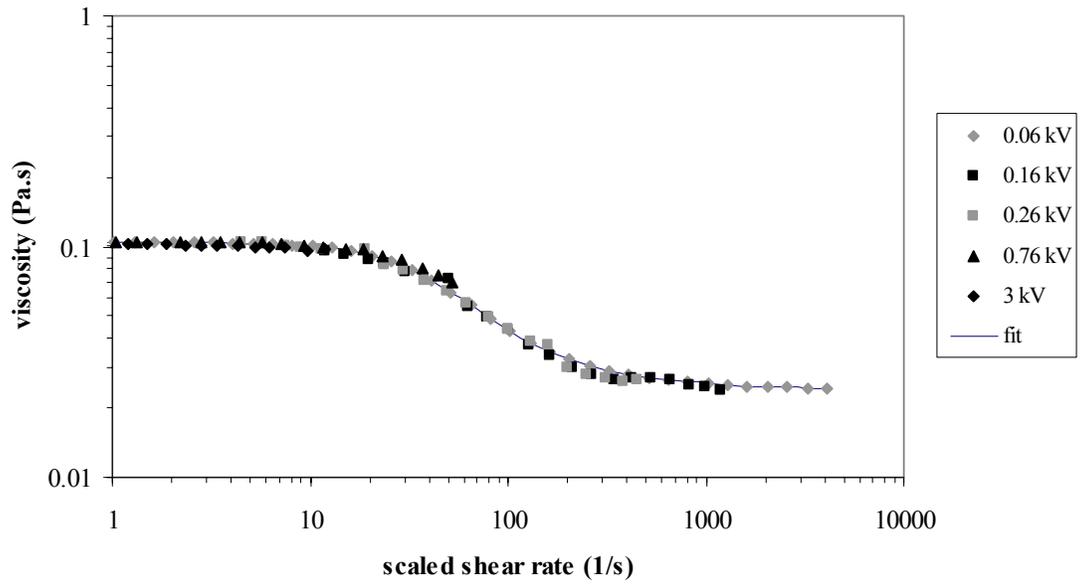

Figure 6

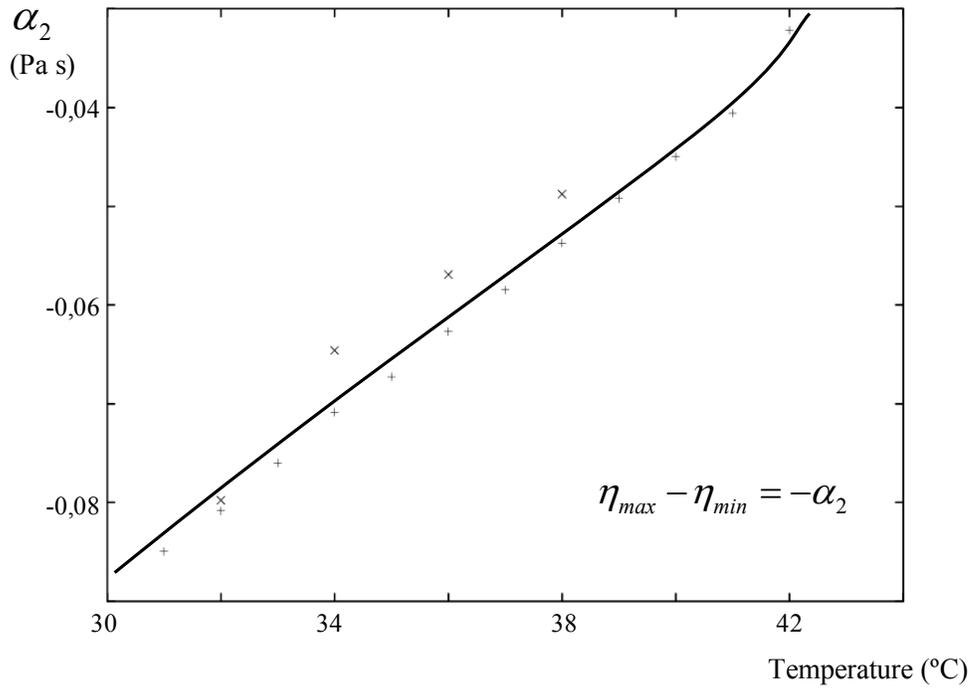

(a)

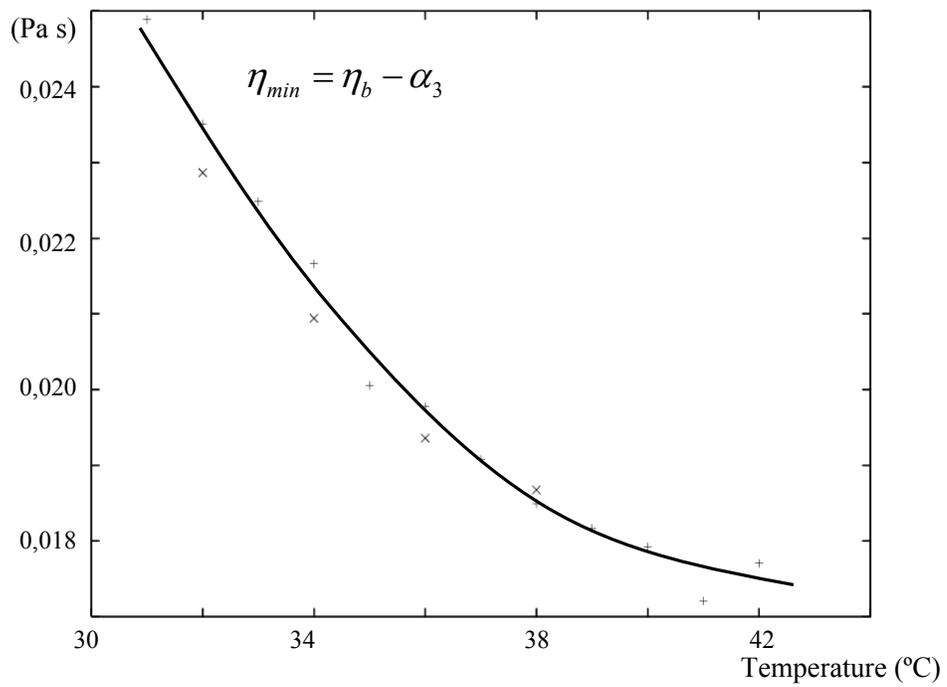

(b)

Figure 7

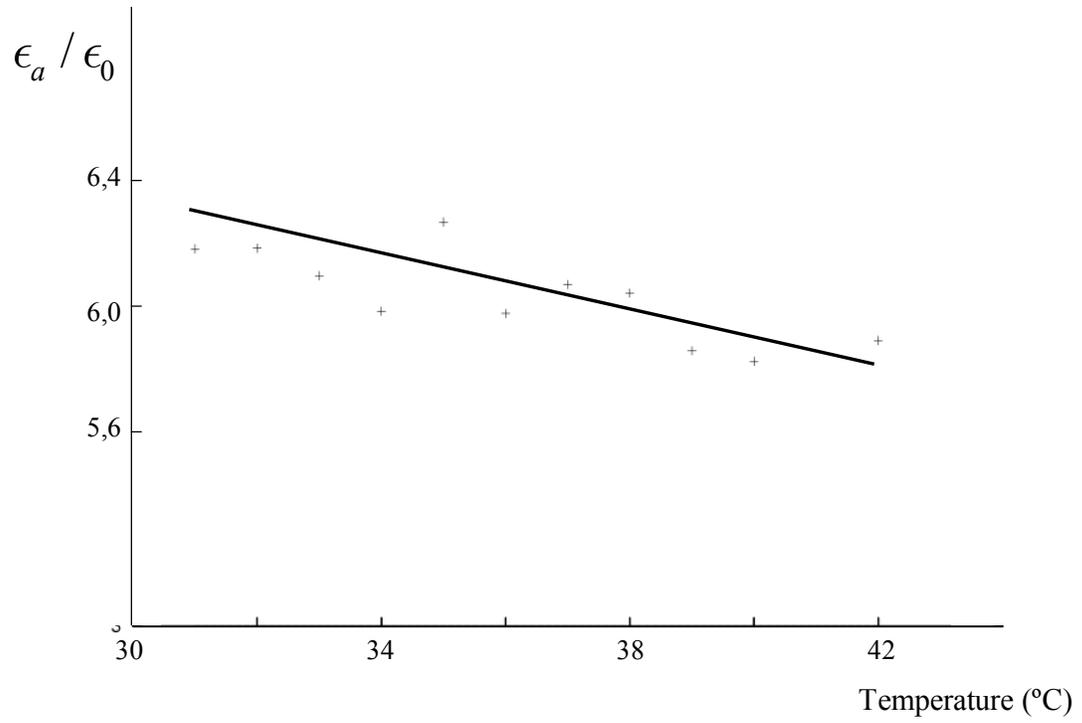